\documentclass[%
 reprint,
unsortedaddress,
 amsmath,amssymb,
 aps,
 onecolumn,
 pra,
longbibliography,
]{revtex4-1}

\usepackage{graphicx}
\usepackage{dcolumn}
\usepackage{bm}

\usepackage[
scale=0.7, marginratio={1:1, 2:3}, ignoreall,
]{geometry}

\usepackage[pdfstartview=FitH,
            breaklinks=true,
            bookmarksopen=true,
            bookmarksnumbered=true,
            colorlinks=true,
            linkcolor=black,
            citecolor=black,
            urlcolor=black
            ]{hyperref}
\usepackage{color}

\newcommand{\subfigure}[1]{(#1)} 
\newcommand{\subfigref}[1]{(#1)} 

\begin{document}

\title{Interaction and breakup of droplet pairs in a microchannel Y-junction}

\author{Simon S. Sch\"{u}tz}
\email{simon.schuetz@epfl.ch}
\affiliation{
Emergent Complexity in Physical Systems Laboratory (ECPS), \'Ecole Polytechnique F\'ed\'erale de Lausanne (EPFL)\\
Station 9, 1015 Lausanne, Switzerland
}

\author{Jian Wei Khor}
\email{jkhor@stanford.edu}
\affiliation{
Department of Mechanical Engineering, Stanford University\\
Stanford, CA 94305, USA
}

\author{Sindy K. Y. Tang}
\email{sindy@stanford.edu}
\affiliation{
Department of Mechanical Engineering, Stanford University\\
Stanford, CA 94305, USA
}

\author{Tobias M. Schneider}
\email{tobias.schneider@epfl.ch}
\affiliation{
Emergent Complexity in Physical Systems Laboratory (ECPS),\\
\'Ecole Polytechnique F\'ed\'erale de Lausanne (EPFL)\\
Station 9, 1015 Lausanne, Switzerland
}

\date{\today{}}

\begin{abstract}
We combine theory, numerical simulation and experiments to investigate the breakup of two identical droplets entering a Y-junction with controlled spatial offset by which the second droplet trails the first.
Based on fully resolved 3D simulations, we describe the flow physics leading to breakup.
Scaling arguments, numerical simulation and experiments consistently show that for small initial offset, breakup always occurs with the droplet fragment volume depending linearly on the offset.
Above a critical offset, which increases with the capillary number, the droplets enter the constriction sequentially without breakup.
Our results are relevant for understanding the breakup behavior in a dense emulsion flowing through a linearly converging channel leading to a constriction.
Such geometry is commonly used for the serial interrogation of droplet content in droplet microfluidic applications, where droplet breakup can limit the maximum throughput for such process.
For capillary numbers up to $Ca\simeq10^{-2}$, the results from the two-droplet system in a Y-junction are consistent with breakup observations in dense emulsions flowing through a linearly converging channel.
The deterministic relation between initial offset and resulting breakup in the two-droplet system suggests that the stochasticity that is observed in the breakup of a dense emulsion arises from multi-droplet interactions.
The numerical value of the prefactor in the linear relation between initial offset and droplet fragment volume determined from experiments differs slightly from the one extracted from fully resolved numerical simulations.
This discrepancy suggests that even at very high bulk surfactant concentrations, the rate-limiting surfactant adsorption kinetics allows for Marangoni stresses to develop and modify the droplet dynamics.
\end{abstract}

\maketitle

\section{Introduction}

In recent years, droplet microfluidics has become a standard tool for high-throughput biochemical assays, including polymerase chain reaction (PCR) \cite{Kiss2008}, in vitro enzyme evolution, and drug screening \cite{Brouzes2009,Agresti2010,Baret2009}.
One of the strengths of droplet microfluidics is the high rate at which droplets can be generated, interrogated, and sorted.
However, this rate can be limited due to the occurrence of droplet breakup at high flow velocities, in particular when droplets are processed as a concentrated or dense emulsion in the microfluidic system.

In dense emulsions flowing in a confined microfluidic system, the occurrence of breakup strongly depends on the microchannel geometry.
Here we focus on a convergent channel that leads to a constriction that fits only one drop at a time, as this geometry is commonly used for the serial interrogation of drops \cite{Rosenfeld2014,Gai2016,Khor2017}.
In this system, the occurrence of breakup appears to be random and the size distribution of breakup fragments reveals chaotic dynamics.
First results on the statistics of droplet breakup in dense emulsions were reported by Rosenfeld et al. \cite{Rosenfeld2014}, who gave the probability of droplet breakup in terms of the capillary number $Ca$, which is the ratio of viscous and surface tension forces.
Gai et al. \cite{Gai2016} reported on the change of breakup probability when varying the droplet size and the viscosity ratio between droplet and continuous phase.
Recently, we observed that for capillary numbers in an intermediate range ($Ca\sim10^{-3}$), droplet breakup depends on the initial offset between the leading edges or the fronts of two droplets entering the constriction (Khor et al. \cite{Khor2017}). 
This observation suggests that droplet breakup in dense emulsions is controlled by the deterministic interaction of two droplets in the constriction, with randomness resulting from the irregular and time-dependent arrangement of droplet pairs in the dense emulsion.

Past research on the breakup of droplets has focused on single droplets, either due to an applied shear or extension from an external flow \cite{Stone1989,Stone1989a} or due to the interaction with microchannel walls at junctions \cite{Link2004,Menetrier2006,Leshansky2009}.
There, droplet breakup occurs when the exterior flow stretches the droplet, forming a neck that undergoes an autonomous pinch-off process \cite{Hoang2013}.
A similar process of induced neck formation and pinch-off has been observed experimentally for the interaction of two droplets in a T-junction \cite{Christopher2009}.

Droplet breakup is controlled by a competition between viscous stress and surface tension. Viscous stress scales with $f_\textrm{visc}=\frac{\mu_d U}{H}$ and promotes the elongation and breakup of the droplet. Surface tension scales with $f_\textrm{surf}=\frac{\gamma}{R}$ and counteracts deformation. $\mu_d$, $U$, $H$, $\gamma$, and $R$ are the droplet viscosity, characteristic droplet speed, channel height, interfacial tension, and undeformed droplet radius, respectively.
The ratio between viscous stress and surface tension,
\begin{equation}
\frac{f_\textrm{visc}}{f_\textrm{surf}} ~=~ \frac{\mu_d U R}{\gamma H} ~\sim~ Ca\cdot \lambda\cdot a
\end{equation}
(with viscosity ratio $\lambda=\frac{\mu_d}{\mu}$, capillary number $Ca=\frac{\mu U}{\gamma}$, and relative droplet size or confinement factor $a=2R/H$), gives an approximate scaling of the onset of droplet breakup \cite{Gai2016}.

In a dense emulsion, the local stresses on a droplet generally result from non-trivial interactions with multiple other droplets.
However, in the geometry of interest, most of the breakup events occur when two primary droplets enter the constriction simultaneously \cite{Khor2017}.
Instead of a dense emulsion, we thus investigate the more tractable problem of two interacting droplets, to gain insights into the physical mechanisms that drive the breakup.

\begin{figure}[!bt]
\centering
\includegraphics[width=\textwidth]{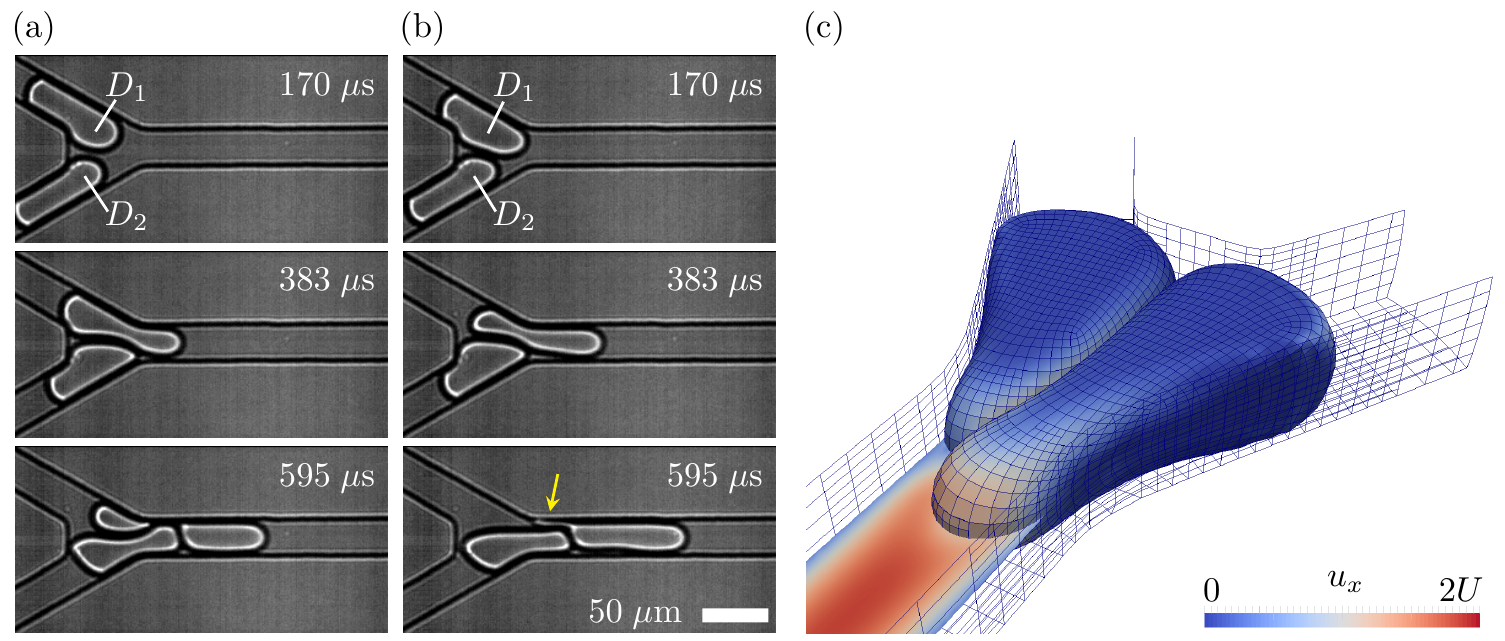}
\caption{
\subfigure{a}, \subfigure{b} Interaction of droplet pairs (from the experiment at $Ca=0.021$). Depending on their initial offset, droplet $D_1$ either breaks up \subfigref{a}, or not \subfigref{b}. The arrow in the last frame of \subfigref{b} points out the shedding of a thin sheet of fluid from droplet $D_1$.
\subfigure{c} Three-dimensional, fully resolved droplet shape (from the simulation at $\alpha=30^\circ$, $Ca=0.06$). The surface mesh has approximately 7,000 vertices, corresponding to 21,000 degrees of freedom. Droplets are colored by the streamwise velocity $u_x$. At mid-height in the channel, we determine the pressure and velocity fields.}
\label{fig:general_geometry}
\end{figure}

In this paper, we report an experimental, numerical and theoretical investigation of the two-droplet interaction that leads to droplet breakup.
We study an isolated system of two identical droplets meeting in a Y-junction (Figure \ref{fig:general_geometry}), where droplet breakup depends on the precisely controlled symmetry-breaking offset between the leading edge of the droplets.
The geometry of the Y-junction is chosen to closely mimick the constriction region in the channel studied by Rosenfeld et al. \cite{Rosenfeld2014} while ensuring that only two droplets enter the junction simultaneously.
The shape of the outer constricting walls is chosen as exactly identical to the channel studied by Rosenfeld et al. Only the forcing behind two interacting droplets differs in that the Y-junction excludes effects due to three or more droplets interacting. The Y-junction thereby allows to study the interaction of two droplets in isolation.
Based on fully-resolved 3D numerical simulation data, we quantitatively describe the physical processes that lead to the breakup process. 
A scaling analysis, fully resolved 3D simulations and precision experiments consistently show that (1) droplet breakup occurs when their leading edges or fronts are below a critical offset, (2) the volume of the fragment of the leading droplet depends linearly on the offset, and (3) the value of the critical offset itself grows with the capillary number so that at higher $Ca$, a wider range of initial offsets leads to droplet breakup.

Precise quantitative comparison of prefactors between experiment and numerical simulation suggests that even at very high bulk surfactant concentrations, non-equilibrium surfactant distributions result in Marangoni stresses that modify the droplet dynamics. 

\section{Problem formulation and methods}
    
\subsection{Problem formulation}
Two identical droplets of volume $V$ and viscosity $\mu_d$ are embedded in a continuous phase of viscosity $\mu$. The droplets enter a Y-junction through two symmetric inlet channels of height $H$ and width $W$ and leave the constriction through a single outlet channel of identical size. The two inlets each join the junction at angle $\alpha$. The total volume flow rate $Q$ is equally split between both inlets. We break the symmetry between both inlets by requiring one droplet to enter the constriction ahead of the other by a leading-edge offset $\delta_0$.

We choose the system height $H$ as our characteristic length scale and the mean outlet velocity $U:=Q/(WH)$ as characteristic velocity scale, which defines a characteristic time scale $\tau:=H/U$.
For a fixed channel geometry, the system is described by three dimensionless parameters, which are the droplet confinement factor
$a := {\sqrt[3]{\frac{6V}{\pi}}}/{H}$, the viscosity ratio $\lambda := \frac{\mu_d}{\mu}$, and the capillary number $Ca := \frac{\mu U}{\gamma}$, where $\gamma$ is the surface tension between the droplet and continuous phase. 
In our system, the confinement factor and viscosity ratio are fixed to $a=1.828$ and $\lambda=0.8$.
At the considered length scales, the effect of inertia can be neglected.
For fixed channel geometry, droplet size and fluids, we have one control parameter, the capillary number (controlled by the flow rate), which we vary between $Ca=0.007$ and $Ca=0.1$.
This is the $Ca$ range where the transition between no breakup and breakup occurs.
The dynamics of droplet breakup depend on the capillary number $Ca$, and the initial droplet configuration, measured by the offset $\delta_0$. 

\subsection{Numerical simulation}
Numerical simulations are performed with a Boundary Element Method (BEM) numerical code based on the \textsc{Deal.II} numerical framework \cite{Bangerth2007}. The \textsc{MPI}-parallel \texttt{C++} code solves the Stokes equations for incompressible Newtonian flow,
\begin{eqnarray}
\mu \nabla^2\vec{u} - \nabla p = \vec{0},\\
\nabla\cdot\vec{u} = 0,
\end{eqnarray}
both in the continuous phase and inside the droplets (with $\mu$ the dynamic viscosity inside the respective domain), under no-slip boundary conditions on the channel side walls, a prescribed velocity profile at the channel inlet, a constant reference pressure at the channel outlet, and a Young-Laplace surface stress of the form
\begin{equation}
\Delta\vec{f} = -2\gamma\kappa\vec{n}
\end{equation}
at the droplet interface, where $\gamma$ is the surface tension, $\kappa$ the mean curvature and $\vec{n}$ the interface normal.
Surfaces are represented by a dynamically refined quad mesh with a second-order (paraboloid) surface shape interpolation.
Time stepping uses a first-order explicit scheme.
The linear system of approximately 21,000 degrees of freedom (for velocity and stress at all interfaces) is solved with the iterative GMRES algorithm.
The mesh is modified after each time step to fix the droplet volumes to the desired value, ensure a minimum gap width between all interfaces of $w_\text{min}=10^{-2}H$, and suppress distortions of the mesh cells.
The code has been validated and accurately reproduces the known flow profile in a rectangular duct, the equilibrium deformation of a droplet in extensional flow and the shape of a droplet flowing through a cylindrical capillary. Details on the implementation and validation of the numerical code can be found in \cite{Schutz2018b}. The negligible influence of the enforced minimum gap width $w_\text{min}$ is discussed in Appendix \ref{sec:appendix_gap_width}.

Simulations are performed for $0.03\leq Ca\leq 0.1$ in increments of $0.01$, for initial droplet offsets $0.05H\leq\delta_0\leq1.5H$.
No breakup is observed for $Ca<0.05$.
The viscosity ratio is $\lambda=0.8$ and the confinement factor is $a=1.828$.
The BEM-scheme does not capture the topological transition of droplet breakup.
Instead, we detect the formation of a neck, and terminate the simulation when the neck width is below $10\%$ of the channel width.
Each simulation run requires between 600 and 2,700 CPU core-hours on a state-of-the-art x86 processor, with simulations at lower $Ca$ demanding more computational effort.
In total, we perform 122 simulations for the different values of $Ca$ and $\delta_0$ mentioned above.

\subsection{Laboratory experiments}\label{sec:methods_exp}
\begin{figure}[!bt]
\centering
\includegraphics[width=\textwidth]{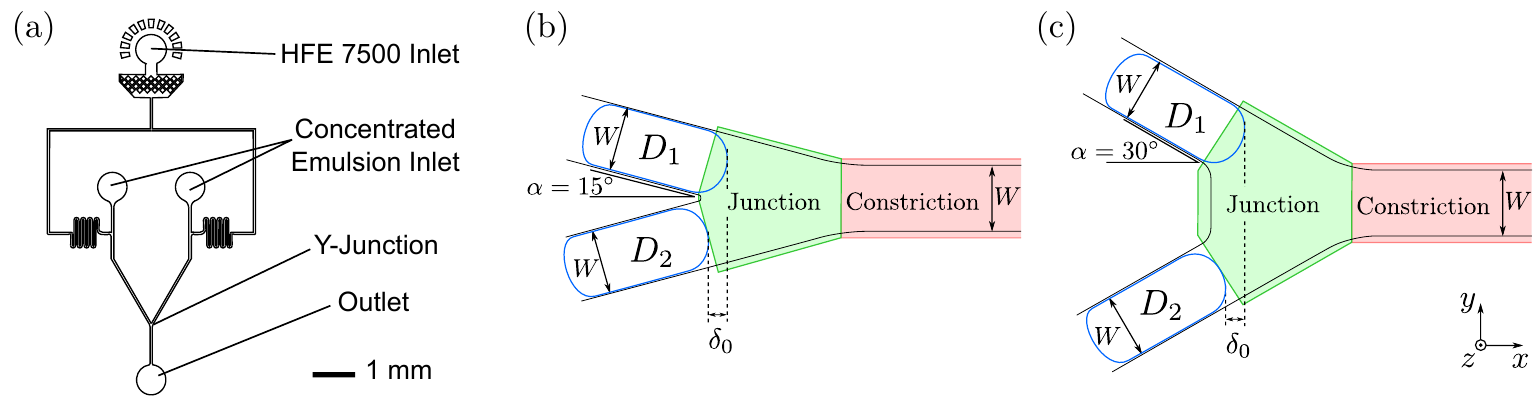}
\caption{
\subfigure{a} Design of the experimental microfluidic device. Microchannels have a height of $H=25\,\mu\textrm{m}$ and a width of $W=30\,\mu\textrm{m}$, flow rates vary from $0.4$ mL/h to $2.0$ mL/h.
\subfigure{b}, \subfigure{c} Y-junction geometry: Two rectangular channels of height $H$ and width $W$ meet at different angles \subfigref{b} $\alpha=15^\circ$, \subfigref{c} $\alpha=30^\circ$. The junction leads into a constriction of width $W$. Two identical droplets $D_1$, $D_2$ arrive at the junction with an offset $\delta_0$ in the streamwise direction.
}
\label{fig:device_design}
\end{figure}

Laboratory experiments use microchannels fabricated in poly(dimethyl\-siloxane) (PDMS) by soft lithography (Figure \ref{fig:device_design}a). The microchannels are bonded to a glass substrate using oxygen plasma and then treated with Aquapel (Pittsburgh, PA) to render the walls of the channel hydrophobic.
The height $H$ of the channels is $25\,\mu\textrm{m}$ and the width $W$ of the constriction channel is $30\,\mu\textrm{m}$. The entrance angle $\alpha$ to the constriction is $15^\circ$ and $30^\circ$ respectively (Figure \ref{fig:device_design}b,c).

We use a flow-focusing nozzle to generate monodisperse droplets. The disperse phase consists of deionized water and the continuous phase consists of HFE-7500 (3M) containing an ammonium salt of Krytox ($2\%$ w/w) as droplet stabilizer.
The interfacial tension between the two liquids is measured to be $26.25$ mN/m using a pendant drop goniometer.
The viscosity of the continuous phase is $1.24$ mPa\,s.
The mean size of the droplets is $50$ pL and the coefficient of variation of droplets is about $3\%$ in volume.
The generated droplets are collected and stored in a 3-mL syringe for $4$ hours at room temperature.
During this time, the drops cream to the top of the syringe to form a dense emulsion, as the drops are less dense than the continuous phase.
The volume fraction of the emulsion, directly measured by counting the number of droplets of known volume $V$ within a given area $A$ of a channel of known height $H$, is about $85\%$. The size of the droplets remains unchanged after their concentration. 

For the break-up experiments, the dense emulsion is split into two syringes. The two syringes of emulsions are injected into a new device via two separate inlets. Immediately downstream from the inlets for the emulsion, extra continuous phase is introduced at $90^\circ$ to both of the branches to lower the volume fraction of the emulsion. The flow rate ratio between the dense emulsion and the continuous phase varies between $0.60$ and $0.67$ to obtain a final droplet volume fraction of about $25\%$ to $35\%$. At this volume fraction, the drops are spaced by a sufficiently large distance from each other to avoid droplet interactions within the same branch prior to entering the constriction. The diluted drops from the two branches then travel downstream and meet at the Y-junction leading to the constriction, where break-up events are recorded. The height of the channel is smaller than the diameter of the droplets when spherical, and the drops always span the whole height of the channel. The total flow rates for our experiment varies from $0.4$ mL/h to $2.0$ mL/h, and are controlled by three syringe pumps (Kent Scientific), two for the dense emulsions, and one for the extra continuous phase to dilute the emulsions. 
The flow rates determine the characteristic time scale $\tau=H/U$ ranging from $30\,\mu\textrm{s}$ to $170\,\mu\textrm{s}$.

An inverted microscope mounted with a high-speed camera is used to acquire images of droplet pairs flowing through the constriction at a frame rate of 45,000 frames per second. This frame rate is sufficiently fast to resolve the leading edges or the fronts of the droplet pairs at the flow rates tested. A custom Matlab code is used to track the location, area, and shape of all droplet pairs ($n > 30$ for each value of $Ca$ and $\alpha$) as well as their broken fragments, and also to measure the offset between the droplet pairs in each frame. Details of the Matlab droplet pair detection algorithm are described in \cite{Gai2016} and \cite{Khor2017}. 

\section{Results and Discussion}
Droplet interaction in the Y-junction proceeds as follows. 
As the two droplets enter the space of the junction, they are separated by a thin vertical oil film.
The front of the first droplet moves into the constriction more quickly, whereas the front of the second droplet slows down.
When the front of the second droplet approaches the constriction, a neck forms in the first droplet.
In situations where the initial leading-edge offset is small, this neck gets progressively thinner and finally pinches off, and the first droplet breaks into two fragments (Figure \ref{fig:general_geometry}a).
Breakup is avoided when the initial offset is sufficiently large (Figure \ref{fig:general_geometry}b).
Similar to the case where two droplets have a small offset, the droplets are separated by a thin oil film, and the front of the first droplet moves faster while that of the second slows down.
However, the rear of the first droplet clears the constriction entrance before a neck can pinch off, and both droplets stay intact.

\subsection{Physical description of the breakup process}
To gain a quantitative understanding of the breakup process, we perform a full 3D simulation of the flow and the droplet interaction in a $30^\circ$-junction at $Ca=0.06$ and offset $\delta_0=0.1H$.
The droplet interface shape is resolved on a sub-$\mu$m scale with a dynamic mesh of approximately 7,000 vertices (Figure \ref{fig:general_geometry}c).
The simulation gives access to the time-dependent full 3D geometry, velocity fields and the pressure.

As the droplets move into the common space of the junction, the front cap of the first droplet $D_1$ maintains a larger radius than the front of droplet $D_2$ due to the initial offset.
Across our simulations, we find that for small but finite offsets ($\delta_0 < 0.3H$) the difference in front radius $\Delta R$ scales approximately with $\Delta R \approx 0.25\delta_0$.

\begin{figure}
\includegraphics[width=\textwidth]{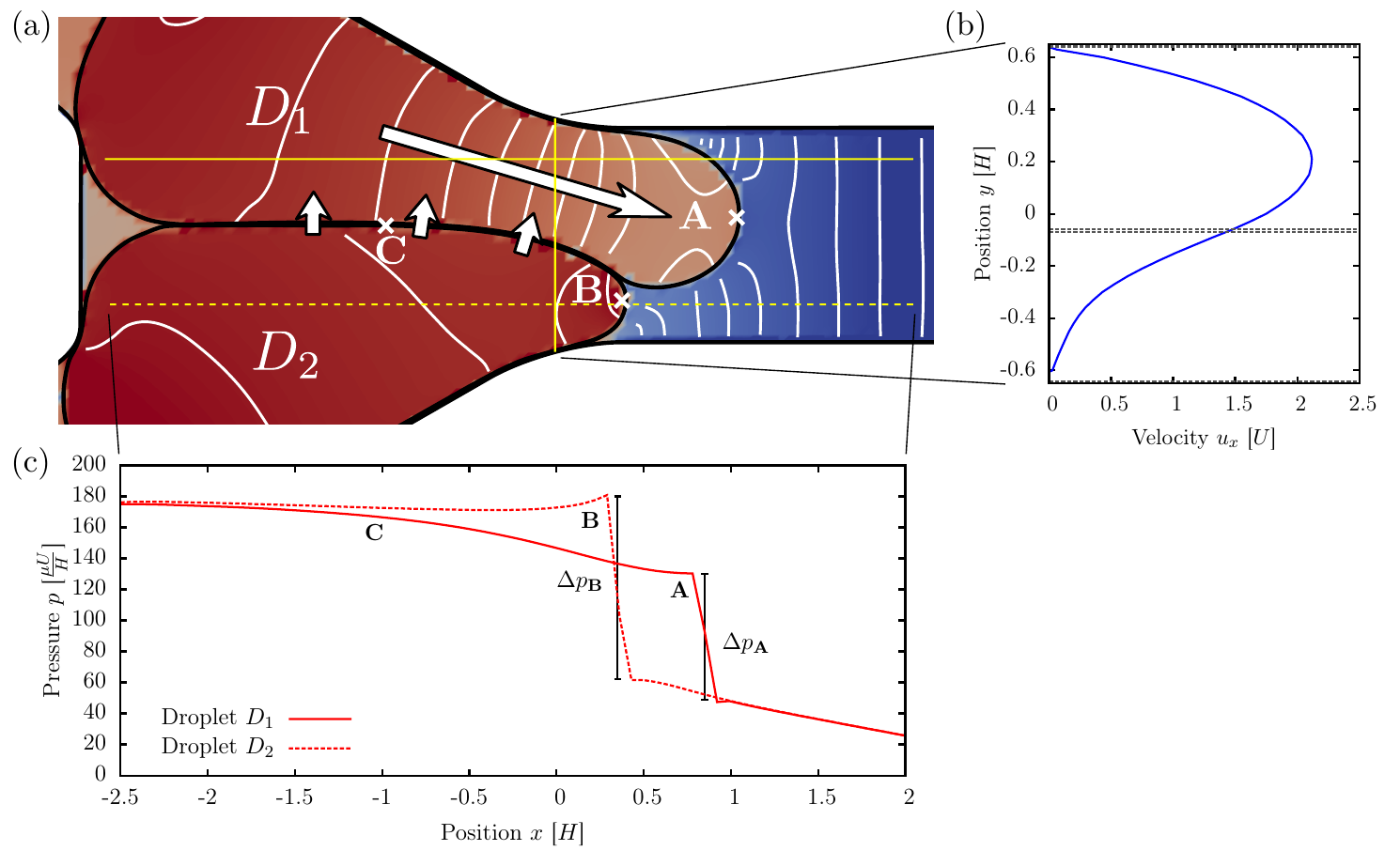}
\caption{Interaction of two droplets with offset $\delta_0=0.1H$ at junction angle $\alpha=30^\circ$ and capillary number $Ca=0.06$ (simulation).
\subfigure{a} Midplane pressure field and contours.
\subfigure{b} Flow velocity at the constriction (vertical solid line in a) in units of the mean flow rate $U$. The dotted line marks the position of the film separating the droplets.
\subfigure{c} Streamwise pressure profile (horizontal full and dashed lines in \subfigref{a}) inside droplets $D_1$ and $D_2$. Due to the different front curvatures at points \textbf{A} and \textbf{B}, the pressure jump $\Delta p_\textbf{A}$ is smaller than $\Delta p_\textbf{B}$, inducing a pressure gradient that drives a relative flow.}
\label{fig:pressure_contour}
\end{figure}

The difference in front radius is the driving force behind the subsequent dynamics. 
For an interface under surface tension, a curvature of the interface leads to a pressure jump between the two sides of the interface, as described by the Young-Laplace equation \cite{Batchelor1967a}
\begin{equation}
\Delta p=\gamma\left(\frac{1}{R_1}+\frac{1}{R_2}\right), \label{eq:YoungLaplace}
\end{equation}
where $R_1$ and $R_2$ are the principal radii of curvature.
The resulting pressure field across both droplets is shown in Figure \ref{fig:pressure_contour}:
When the droplets reach the constriction, the horizontal front radius of $D_1$ (point \textbf{A}) is larger than that of $D_2$ (point \textbf{B}), so that the pressure jump $\Delta p_\textbf{A}$ is smaller than the pressure jump $\Delta p_\textbf{B}$.
As the pressures are similar towards the back of the droplets, the pressure gradient within the bulk of $D_1$ (\textbf{C} $\rightarrow$ \textbf{A}) is larger than inside $D_2$ (\textbf{C} $\rightarrow$ \textbf{B}).
The difference in pressure gradient drives a relative flow between the droplets (Figure \ref{fig:pressure_contour}c).
The relative flow increases the offset $\delta$ and thus the front radius of $D_1$, the process is self-reinforcing and leads to drainage of $D_1$ ahead of $D_2$, where drainage describes the transport of fluid volume through the constricted neck. The drainage rate is proportional to the surface tension $\gamma$ and inversely proportional to the shear stress $\mu_d U$ that counteracts the flow. 
A quantitative estimate of the drainage rate is given in Appendix \ref{sec:appendix_drainage_rate}.

\begin{figure}
\includegraphics[width=\textwidth]{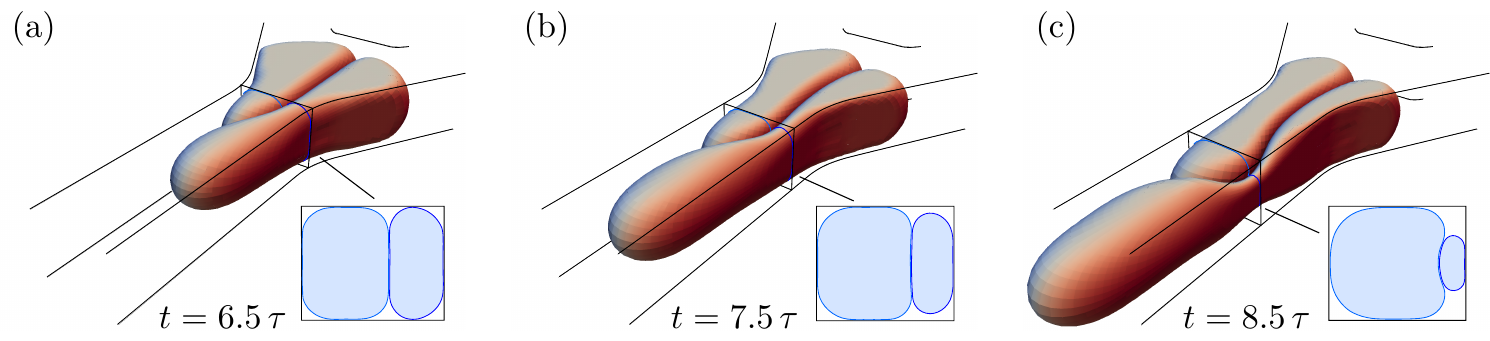}
\caption{Interaction of two droplets with offset $\delta_0=0.1H$ at junction angle $\alpha=30^\circ$ and capillary number $Ca=0.06$, at different times $t$ after droplets enter the junction. Images from the simulation. A neck forms in the leading droplet, first only in the horizontal direction \subfigref{a}, then also vertically \subfigref{b}. After the neck starts forming in the vertical direction, pinch-off happens quickly \subfigref{c}.}
\label{fig:necking}
\end{figure}

The formation of a neck and subsequent pinch-off interrupt the drainage process.
Along the water/oil/water interface behind the front caps, the pressure in droplet $D_2$ is higher than in droplet $D_1$.
This pressure difference is not fully compensated by the curvature of the interface between the droplets, and leads to a cross-stream motion of the interface towards droplet $D_1$, so that droplet $D_1$ elongates and a neck forms.
As the neck enters the constriction, it gets so narrow that curvature in the vertical $z$-direction becomes very large (Figure \ref{fig:necking}).
The Young-Laplace pressure jump due to the large curvature in the vertical direction is not supported by the pressure inside the droplet, so that the interface of $D_1$ separates from the top and bottom channel walls.
From the point where the neck starts to cave in vertically, the process is self-reinforcing and resembles the Rayleigh-Plateau instability \cite{Plateau1873,Eggers1997a}.
For a similar case of single-droplet breakup, Leshansky \cite{Leshansky2009} and Hoang \cite{Hoang2013} found the transition to this autonomous pinch-off to lie at a neck width of $0.5H$ in channels with near-unity aspect ratio.

The time scale $\Delta T$ for the breakup process is dominated by the advection time between when the droplets meet in the junction, and when the droplet caps arrive in the constriction. 
Therefore, this time depends on the junction geometry and characteristic time scale $\tau\equiv H/U$.
Relative to the advection time scale, the formation of the vertical neck and pinch-off happen quickly.
Breakup is avoided when the entire volume of $D_1$ can drain ahead of the necking region during the time $\Delta T$.

In summary, droplet interaction in a Y-junction is dominated by a drainage flow driven by surface tension.
The drainage, by which one droplet moves ahead of the other, is interrupted by the formation of a neck due to a difference in internal pressure between the droplets.
In the constriction, the neck formation becomes self-reinforcing, and leads to pinch-off.

\subsection{Dependence of the droplet fragment volumes on the initial offset}

\begin{figure}[!b]
\centering
\includegraphics[width=0.7\textwidth]{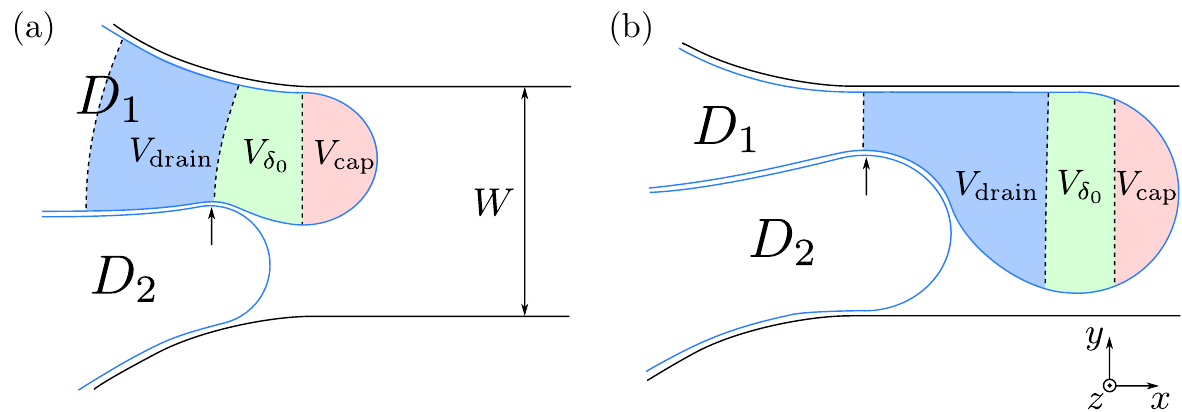}
\caption{Sketch of the drainage process (not to scale). The final volume $V_{1a}$ of the first fragment of $D_1$ after pinch-off comprises three parts: The volumes $V_{\delta_0}$ and $V_\textrm{cap}$, which are ahead of the neck before the start of necking \subfigref{a}, and the volume $V_\textrm{drain}$ that drains through the neck before it pinches off completely \subfigref{b}. The neck forms behind the front cap of droplet $D_2$ (arrow).}
\label{fig:fragmentscaling}
\end{figure}

We now use our understanding of the physical processes that drive droplet breakup to model the dependence on capillary number and initial droplet offset.
In particular, we describe the effect of these parameters on the volume of the breakup fragments, which we can quantitatively determine in both simulation and experiment.
With $V_1$ the volume of droplet $D_1$ before entering the junction, we consider the volume $V_{1a}$ of the first fragment of this droplet after passing the constriction.
If $V_{1a}=V_1$, no breakup has occurred. If $V_{1a}<V_1$, breakup has occurred, and at least two fragments have been created.

Neck formation takes place behind the front cap of droplet $D_2$, and only after some time $\Delta T$, during which the fluid in $D_1$ drains ahead of the neck.
The volume $V_{1a}$ is then made up of three distinct parts.
These parts, as illustrated in Figure \ref{fig:fragmentscaling}, are the volume $V_\textrm{cap}$ of the front cap of $D_1$, the volume $V_{\delta_0}$ by which the first droplet was ahead of the second as they entered the junction, and the volume $V_\textrm{drain}$ that drains ahead of the neck during the neck formation.
We approximate the shape of the front cap $D_1$ as a half-ellipsoid of radius $W/4$ in the horizontal plane and half-height $H/2$, with a volume of $V_\textrm{cap}\approx\frac{\pi}{24}W^2H$.
The volume $V_{\delta_0}$ depends on the initial offset $\delta_0$ and the cross-sectional area of the inlet channel, and is approximated as $V_{\delta_0}\approx\frac{WH\delta_0}{\cos\alpha}$.

Since the relation between the offset and the front radius difference $\Delta R$ is unknown, an estimation of the drained volume $V_\textrm{drain}$ proves difficult. However, we know that the drained volume will grow with the initial offset $\delta_0$ (which determines the initial curvature difference) and the factor $\frac{\gamma}{\mu_d U}\equiv\frac{1}{\lambda Ca}$ (which drives the drainage flow based on that curvature difference). This suggests a draining volume of $V_\textrm{drain}\approx C_1\cdot\frac{\delta_0}{\lambda Ca}$, where $C_1$ is an unknown constant.
The sum of the three volume components gives the estimate
\begin{equation}
V_{1a}~=~ V_\textrm{cap}+V_{\delta_0}+V_\textrm{drain} ~\approx~\frac{\pi}{24}W^2H + \frac{WH}{\cos\alpha}\cdot\delta_0 + C_1\cdot\frac{\delta_0}{\lambda Ca}.\label{eq:scalingmodel}
\end{equation}

From this relation, we get several predictions for the scaling of the first fragment volume.
We expect the fragment volume to grow linearly with the offset, and with a steeper slope in the case of smaller capillary numbers.
We define the critical offset $\delta_\textrm{crit}$ as the value of $\delta_0$ at which the fragment volume reaches the full droplet volume.
No breakup occurs for $\delta_0\geq\delta_\textrm{crit}$.
If the linear relation between initial offset and fragment volume is steeper at lower $Ca$, we expect $\delta_\textrm{crit}$ to be small at low $Ca$, and increase as $Ca$ gets larger.

\subsection{Scaling of the fragment volumes: Simulations}
\begin{figure}
\includegraphics[width=\textwidth]{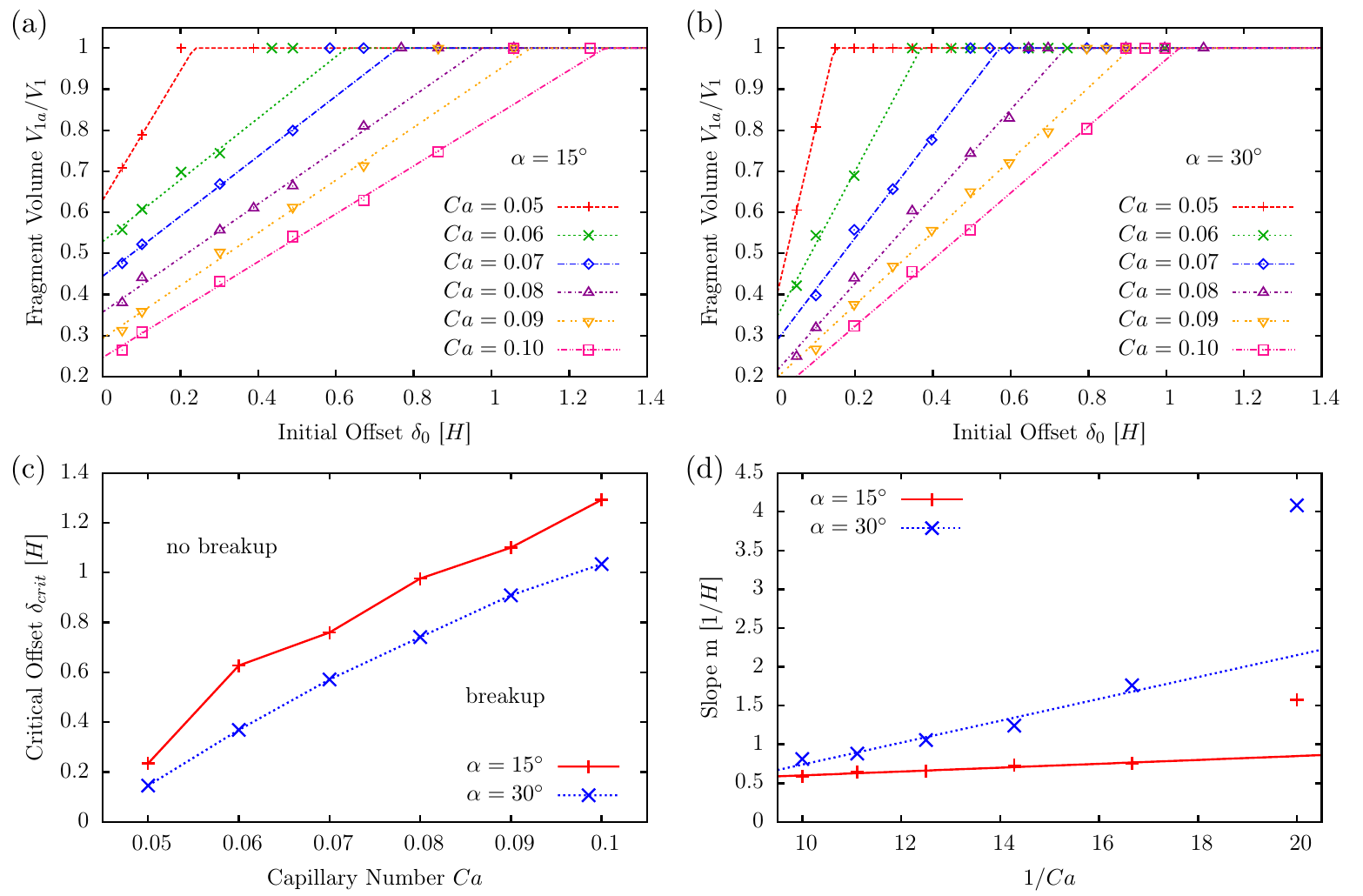}
\caption{Simulation results for the relation between initial droplet offset and the resulting droplet breakup. \subfigure{a}, \subfigure{b} Relative volume of the first droplet fragment after breakup, at a junction angle of $\alpha=15^\circ$ \subfigref{a} and $\alpha=30^\circ$ \subfigref{b}. As the relative volume reaches one, no breakup occurs. Dashed lines show the piecewise linear trend. \subfigure{c} Critical offset at which no breakup occurs, as function of capillary number $Ca$. The critical offset $\delta_{crit}$ is the offset $\delta_0$ at which the linear extrapolation reaches 1. \subfigure{d} Slope $m$ of the linear relation $V_{1a}/V_1=a+m\cdot\delta_0$ between fragment volume $V_{1a}$ and initial offset $\delta_0$ (taken from the linear trend in \subfigref{a}, \subfigref{b}), against the inverse capillary number $1/Ca$. The data match predictions from the scaling model (eq. \ref{eq:scalingmodel}), which predicts a linear relationship with an unknown prefactor $C_1$. For small capillary numbers (large $1/Ca$) the slope is steeper and deviates from the linear scaling.}
\label{fig:volumefraction}
\end{figure}

We test the analytical prediction using extensive numerical simulations with adaptively refined, highly resolved surface meshes (21,000 degrees of freedom) and at high temporal resolution (adaptive time steps around $10^{-3}\tau$) for 122 different combinations of capillary number and initial offset.
The junction angles are $\alpha=15^\circ$ and $\alpha=30^\circ$, the capillary number range is $0.05\leq Ca\leq0.1$ and initial offsets are in the range $0<\delta_0<1.4H$.
The confinement factor is chosen as $a=1.828$.

For the entire range of offsets, the fragment volume $V_{1a}$ displays the linear dependence on the initial offset as predicted from the model (Figure \ref{fig:volumefraction}a).
In accordance with our expectations, the linear relation between initial offset and fragment volume scales with $1/Ca$ (Figure \ref{fig:volumefraction}d).
The critical offset $\delta_\textrm{crit}$ grows with the capillary number (Figure \ref{fig:volumefraction}c).
At small $Ca$, breakup only occurs for highly symmetric droplet configurations with small offset. At large $Ca$ ($Ca=0.1$), we observe breakup in droplet configurations with large offset, even in situations where one droplet is ahead of the other by half of its length.

A larger junction angle decreases the range of offsets where breakup is observed. For junction angle $\alpha=30^\circ$, the values for the critical offset are smaller than for $\alpha=15^\circ$. This is due to the larger droplet width in the junction for large junction angles, which allows for faster drainage flows.

We find that for infinitesimal initial offsets ($\delta_0\approx0$), $V_{1a}$ is larger for smaller values of $Ca$ (Figure \ref{fig:volumefraction}a,b).
Experiments confirm this behavior.
The dependence on $Ca$ for infinitesimal initial offsets is likely a consequence of the variation in droplet shape with the capillary number.
At large $Ca$, the droplet cap is more pointed \cite{Lac2009}, so that the volume of the droplet cap, which determines the fragment volume, is smaller.

While our analysis of two-droplet interaction focuses on one specific droplet size ($a=1.828$), previous experiments on dense emulsions \cite{Rosenfeld2014,Gai2016} show that the droplet interaction and breakup are not sensitive to the droplet size.
This insensitivity to the droplet size is confirmed by simulations of the two-droplet interaction in the Y-junction. 
Decreasing or increasing the droplet volume by 30\% hardly impacts the pinch-off point, but only affects the volume of the upstream droplet fragments (data not shown).
We speculate that upstream droplet fragments might undergo additional pinch-off events if the remaining volume after the initial pinch-off is sufficiently large.

\subsection{Scaling of the fragment volumes: Experiments}
\begin{figure}
\includegraphics[width=\textwidth]{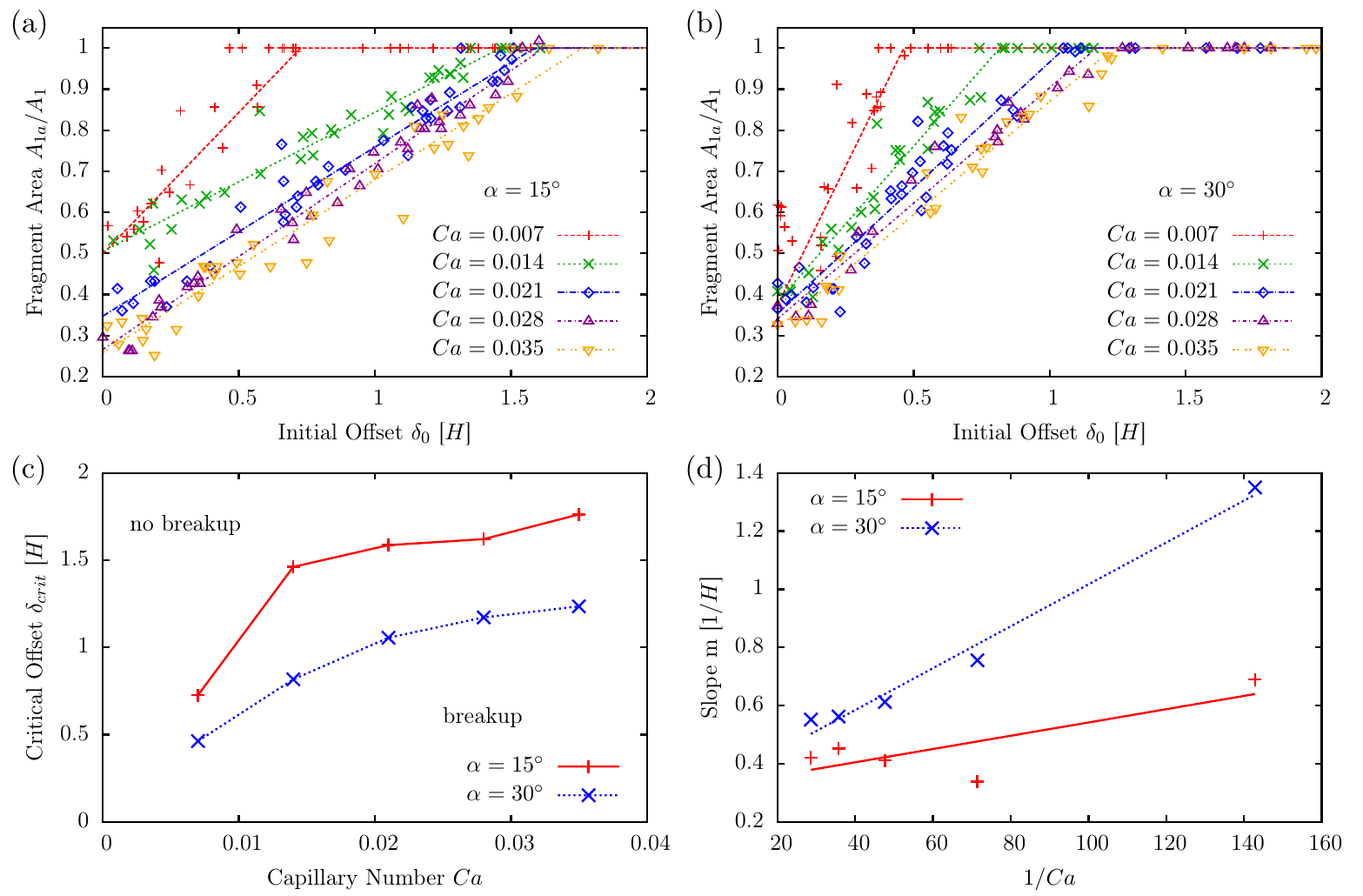}
\caption{Experimental results for the relation between initial droplet offset and the resulting droplet breakup. \subfigure{a}, \subfigure{b} Relative area of the first droplet fragment after breakup. As the relative area reaches one, no breakup occurs. Dashed lines show the piecewise linear trend. \subfigure{c} Critical offset at which no breakup occurs, as function of capillary number.
\subfigure{d} Slope $m$ of the linear relation $A_{1a}/A_1=a+m\cdot\delta_0$ between fragment area $A_{1a}$ and initial offset $\delta_0$ (taken from the linear trend in \subfigref{a}, \subfigref{b}), against the inverse capillary number $1/Ca$. Solid lines show the linear trend predicted by the scaling model (eq. \ref{eq:scalingmodel}).}
\label{fig:experimental}
\end{figure}

Experimental measurements confirm the results of our simulations.
We perform experiments for the same Y-junction geometry as in the simulations, with the same junction angles and droplet size. 
By varying the flow rate, we measure capillary numbers in the range $0.007\leq Ca\leq0.035$.
We do not actively control the offset between the two droplets coming in from the two branches prior to the Y-junction, but take advantage of the random variation in the spacing of droplets.
By examining a large number of droplet pairs, this random variation in spacing conveniently allows us to obtain a large number of initial offset values $\delta_0$ without the need for complicated active flow control.
As the optical setup only allows imaging of the droplets in the horizontal plane, we measure the area of the droplets in this image plane as a measure for their volume.
As seen in Figure \ref{fig:experimental}, while there is a larger variation in the experimental data, they display the same behavior as the simulation results.
The relative size of the first fragment grows linearly with the initial offset, and at a steeper slope for lower capillary numbers (Figure \ref{fig:experimental}a,b).
Consequently, the critical offset for droplet breakup grows with the capillary number, so that for larger capillary numbers, breakup is a common phenomenon, which occurs even for offsets on the scale of the droplet length (Figure \ref{fig:experimental}c).
As in the simulations, a larger junction angle results in a smaller critical offset. 

Both the data from the simulation and from the experiments show the behavior that we expect from the theoretical scaling analysis.
They explain how droplet breakup is more frequently observed at higher capillary numbers.
The drainage flow, which prevents breakup, is driven by surface tension and scales with the inverse of the capillary number.
For small capillary numbers, the critical offset for breakup is small, so that it is statistically unlikely for a droplet pair to have an even smaller offset leading to breakup.
At larger capillary numbers, there is an increased range in offsets for which breakup is possible, so that droplet breakup becomes more and more frequent.

While we find great qualitative agreement between the fragment size data from the simulation and experiment, the experiments reveal an apparent quantitative discrepancy in the capillary number at which breakup is observed.
In the experiments, droplets break up at capillary numbers that are lower than the numerical predictions by a factor of 5--8.
This prefactor is only of order unity, but cannot be attributed to experimental uncertainties or numerical error alone.
In the following we provide quantitative evidence that the discrepancy is due to nonequilibrium surfactant concentrations causing Marangoni stresses that are intentionally not included in the simulation.

\begin{figure}
\centering
\includegraphics[width=\textwidth]{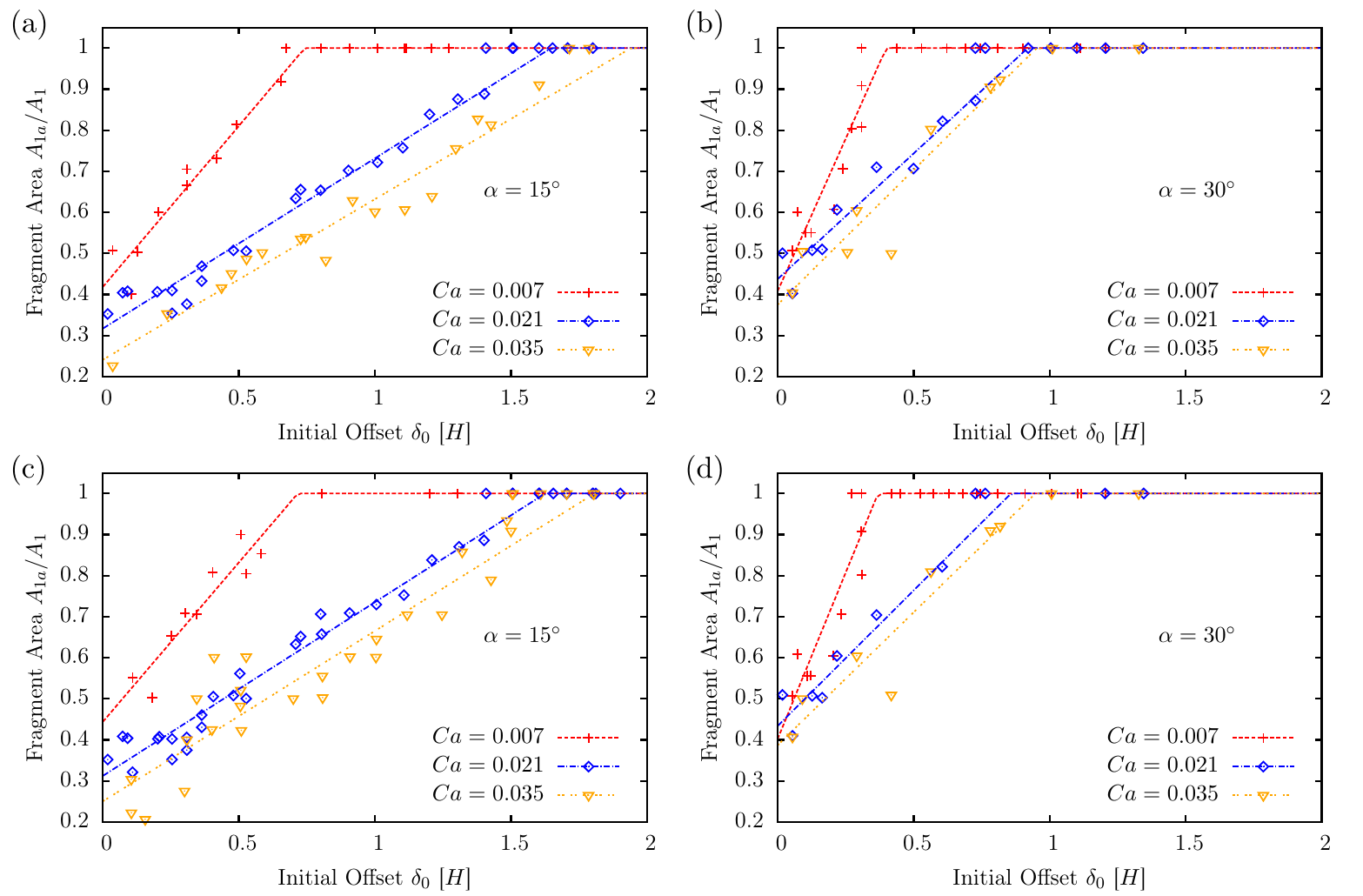}
\caption{
Additional experimental checks with a Reynolds number reduced by a factor of two (\subfigure{a}, \subfigure{b}) and for a surfactant concentration reduced by a factor of 20 (\subfigure{c}, \subfigure{d}) show no significant modification of the relation between initial droplet offset and the resulting droplet breakup compared to Figure \ref{fig:experimental}.\\
\subfigure{a}, \subfigure{b}: Relation between fragment size and initial offset for a scaled down geometry of height $H=12.5\mu\text{m}$ and width $W=15\mu\text{m}$ with opening angle \subfigref{a} $\alpha=15^\circ$, \subfigref{b} $\alpha=30^\circ$, where the droplet volume and flow rates have been adjusted to maintain the shape factor and capillary number.\\
\subfigure{c}, \subfigure{d}: Relation between fragment size and initial offset for a reduced surfactant concentration of $0.1 \%$ w/w, in a geometry with opening angle \subfigref{c} $\alpha=15^\circ$, \subfigref{d} $\alpha=30^\circ$.
}
\label{fig:experimental_variation}
\end{figure}

First, we exclude the potential influence of finite Reynolds number effects, which the simulation neglects, as source of the difference.
At the highest flow rate, $Q=2\;\textrm{mL}/\textrm{h}$ ($Ca=0.035$) and thus highest Reynolds number, the local acceleration of the droplet front is $U/\tau\approx2\cdot10^4\;\textrm{m}/\textrm{s}^2$ for a fluid volume of roughly $H^3=15\,\textrm{pL}$.
The resulting inertial force of $\rho H^3 U/\tau=0.3\,\mu\textrm{N}$ is small compared to the $4\gamma H=2.6\,\mu\textrm{N}$ of the Laplace pressure acting on the droplet cap.
Consequently, finite Reynolds number effects are too small to have a significant effect.
Further evidence that inertial finite Reynolds number effects are negligible is provided by experiments at a reduced Reynolds number. Halving both all length scales and the flow rates reduces the Reynolds number by a factor of two, yet no significant impact on the breakup dynamics is observed (Figure \ref{fig:experimental_variation}a,b).

While inertial effects are shown to be negligible, we propose and provide quantitative evidence that the reason for the discrepancy between experiments and simulations is a nonequilibrium distribution of surfactants on the droplet interfaces, which has not been modeled as part of the simulation.
For the surfactant used in the experiment (Krytox), the adsorption time to the interface is known to be on the scale of tens of milliseconds, and is determined by the kinetics of the adsorption process rather than the bulk concentration \cite{Riechers2016,Baret2009a}. 
We confirm this by observing that reducing the bulk concentration by a factor of $20$ from $2\%$ w/w to $0.1\%$ w/w does not affect the experimental results (Figure \ref{fig:experimental_variation}c,d).
The time scale of droplets passing the constriction, $\Delta T\approx 60-340\;\mu\textrm{s}$, is two orders of magnitude smaller than the adsorption time.
We can thus assume that almost no additional surfactant is adsorbed during the process.
This has two important consequences.
First, since the total droplet area changes during the deformation process, the area concentration of surfactant and thus the surface tension changes accordingly.
Second, the local expansion and contraction of the surface creates Marangoni stresses, which act along surface tension gradients in the interface plane.
These Marangoni stresses counteract the deformation and thus prevent drainage, which promotes droplet breakup even at lower capillary numbers than those encountered in the simulation.

\begin{figure}
\centering
\includegraphics[width=0.5\textwidth]{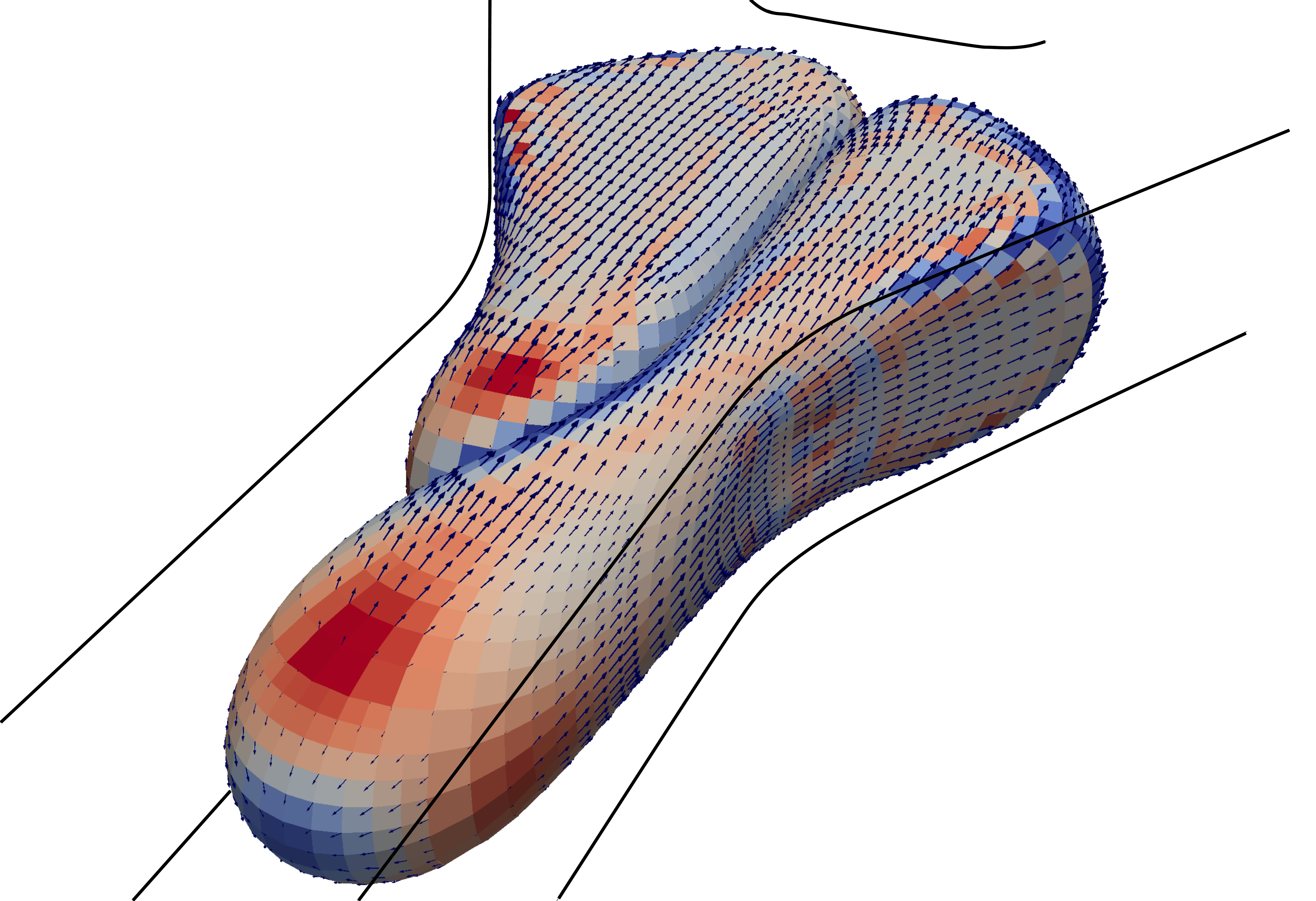}
\caption{In-plane velocity divergence on the droplet surface, at $Ca=0.06$ and $\delta_0=0.1H$. Images from the simulation. Colors show the in-plane divergence of the velocity on the droplet surface, in the range between $-3\,\tau^{-1}$ (blue) and $3\,\tau^{-1}$ (red). Vectors show the surface velocity relative to the mean velocity of $D_1$. The flow sweeps surfactants towards the back of the droplet, causing strong variations in the surfactant density.}
\label{fig:marangoni}
\end{figure}

Quantitatively, it has been found that the surfactant typically decreases surface tension by roughly a factor of two \cite{Riechers2016}, 
so that the depletion of surfactants could effectively double the Laplace pressure in the droplet front cap.
From the simulation, we extract the in-plane flow on the droplet surface, which redistributes surfactants and moves them to the rear of the droplets.
The resulting flow field is shown in Figure \ref{fig:marangoni}, together with a color-coding for regions where the interface expands (red) or get compressed (blue). The total interfacial area of Droplet 1 expands by $32\%$ during the droplet breakup (Droplet 2: $12\%$), with a much higher local expansion in the front cap (red areas in Figure \ref{fig:marangoni}). Even though the exact relation between surfactant density and surface tension is not known in detail, both the Young-Laplace pressure that drives the drainage and the Marangoni stresses that inhibit it are on the order of $\gamma/R$. Consequently, variations in surface tension are strong enough to suppress the drainage mechanism. When Marangoni stresses suppress the drainage, droplets will enter the constriction together even at lower $Ca$. This shifts the onset of observed breakup to lower $Ca$, as observed in the experiments.

\begin{figure}
\centering
\includegraphics[width=\textwidth]{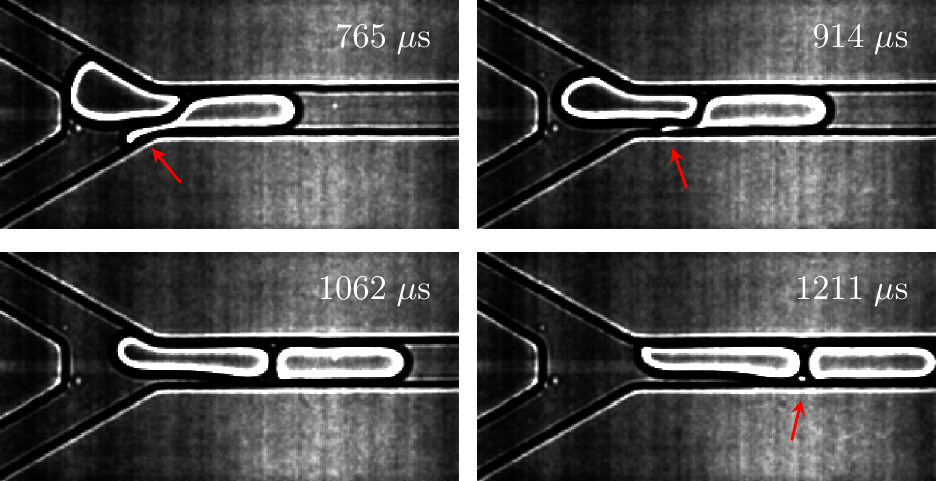}
\caption{
When droplets enter the constriction at high $Ca$, the interface of the leading droplet can be sheared off, forming small satellite droplets. Shown here is the process at $Ca=0.014$, $\lambda=0.8$, in a junction with $\alpha=30^\circ$. Arrows indicate the formation of the satellite droplet, which moves into the space between the two primary droplets but does not coalesce with either of the droplets. Videos of the process are available as supplemental online material \cite{SupplementaryVideo}.
}
\label{fig:satellite_droplets}
\end{figure}

The effect of the nonequilibrium surfactant distribution can be directly observed in the microscope images from the experiment.
If the droplet gets too close to one side wall (Figure \ref{fig:general_geometry}b, Figure \ref{fig:satellite_droplets}), a narrow sheet or finger of the droplet surface is swept away, breaking up into tiny droplets.
We do not classify this behavior as breakup of the leading droplet, as the satellite droplet is very small in size and moves into the gap between the two primary droplets.
The formation of satellite droplets is not observed in the numerical simulations of droplets with Young-Laplace surface tension.
The strong deformation at the rear end of the droplet suggests that surfactants accumulate in the rear stagnation point due to advective bulk fluid motion, and lower the interfacial tension.
If this formation of satellite droplets, by which surfactant is removed from the droplet surface \cite{Stone1994}, occurs also in the upper parts of the channel, it is likely that the surfactant concentration is below the equilibrium concentration even before entering the junction.

\subsection{Implications for dense emulsions}
Our results on two-droplet interactions here are relevant to a dense emulsion entering a convergent channel leading to a constriction as in the latter, droplet breakup results primarily from the interaction of two droplets at the constriction entrance \cite{Khor2017}. However, there are also differences between the two systems. 

In the two-drop system flowing into a Y-junction, the shape, orientation and angle between the two drops are fixed. As such, the interaction between the drops, and ultimately the droplet fate (breakup or not), is a deterministic process as governed by the leading-edge offset between the two drops alone for a given set of flow conditions, channel geometry and droplet properties. 

In contrast, in a dense emulsion system flowing into a converging channel, the presence of other drops in the emulsion adjacent to the two primary drops entering the constriction leads to a range of variabilities in the shape, orientation, and angle between the two primary drops. For example, the presence of a third drop stacked on top of the two primary drops leads to a smaller effective angle at which the two drops approach the constriction. This smaller angle in turn promotes droplet breakup. Such variabilities add a chaotic component to the droplet interaction and fate, which manifests itself as a bistable region where both breakup and non-breakup events are observed even when the offsets between the two primary drops are identical. Such bistable region is absent in the two-drop system flowing into a Y-junction both in our simulations and in our experiments.

When we combine our experimental data on the critical offset for droplet breakup in a Y-junction with the distribution of initial offsets in dense emulsions \cite{Khor2017}, we successfully predict the probability of droplet breakup in dense emulsions at capillary numbers up to $Ca\simeq10^{-2}$ within a $5\%$ error.
The prediction uses the leading-edge offset $\Delta x$ defined in \cite{Khor2017}.
Breakup data for this offset definition is available as supplemental material \cite{SupplementaryOffset}.

\section{Conclusion}
By combining theoretical analysis, numerical simulations and experimental study, we have described the breakup process of two droplets entering a Y-junction.
The droplet interaction follows a two-step process.
In a first step, one droplet moves ahead of the other droplet, driven by an internal pressure gradient due to surface tension.
In a second step, a neck forms in the first droplet, which eventually breaks up in an autonomous pinch-off process.
If the drainage of the rear end of the first droplet completes before the neck pinches off, breakup is avoided.
The strength of the drainage process, and thus the occurrence of breakup, is determined by the offset between the droplets prior to entering the constriction.

A scaling analysis for the volume of the breakup fragments reveals a linear dependence between the volume of the first droplet fragment and the initial leading-edge offset between the droplets.
This linear dependence is verified in both simulations and experiments.

Quantitatively, droplet breakup in the experiment occurs at capillary numbers significantly lower than in the simulation that assumes the absence of Marangoni stresses.
We provide evidence suggesting that the observed difference is caused by nonequilibrium surfactant concentrations, which can exist because the time scale of surfactant adsorption is large compared to the advective time scale of the interface dynamics.
Nonequilibrium gradients in surfactant density cause in-plane Marangoni stresses, which can oppose drainage and promote breakup.
Confirming the hypothesized influence of Marangoni stresses requires a detailed understanding of surfactant adsorption kinetics, which is subject of ongoing research \cite{VanHunsel1986,Stone1990a,Song2006,Baret2009a,Riechers2016,Ponce-Torres2017}.

Using our experimental data on the critical offset for droplet breakup in a Y-junction, we successfully estimate the probability of droplet breakup in dense emulsions. This suggests that the two-droplet interaction studied in this work also dominates breakup in dense emulsions. Since the breakup process is deterministically controlled by the relative offset of two droplets, the stochasticity within dense emulsions appears due to the randomness of relative offsets arising predominantly from multi-droplet interactions.

\section*{Acknowledgement}
JK and ST acknowledge support from the National Science Foundation through the NSF CAREER Award No. 1454542.


%

\clearpage
\appendix

\section{Magnitude of the Relative Flow due to the Difference in Front Radius}\label{sec:appendix_drainage_rate}
For rectangular channels of aspect ratios $H/W$ near unity, the relation between streamwise pressure gradient $\partial_x p$ and approximate mean flow velocity $U$ is
\begin{equation}
U~=~-0.035\,\frac{WH}{\mu}\cdot\partial_x p\label{eq:vel_pressure}
\end{equation}
where $\mu$ is the dynamic viscosity of the fluid inside the channel. This result can be reached by numerically evaluating the series representation of the analytic flow profile in the duct \cite{Spiga1994}.

Consider now the case of the two front caps of the droplets lying next to each other with difference in horizontal front radius $\Delta R\equiv R_1-R_2\ll 1$, each occupying about half of the width of the channel ($R_1\approx R_2\approx W/2$). With the ambient pressure the same between the droplets, the internal pressure will be higher in $D_2$ than $D_1$ by
\begin{equation}
\Delta p_{21} ~=~ \gamma\left(\frac{1}{R_2}+\frac{1}{H/2}\right) - \gamma\left(\frac{1}{R_1}+\frac{1}{H/2}\right) \approx \frac{\gamma \Delta R}{(W/2)^2},
\end{equation}
according to the Young-Laplace equation \eqref{eq:YoungLaplace}, with $\Delta p_{ij}\equiv p_i-p_j$.

This pressure difference exists at the front cap of the droplets, where it is supported by the local curvature of the interface, but not further back, where the water/oil/water interface between the droplets runs straight. This is the case at about a distance $W$ from the droplet fronts, such that each droplet has an internal pressure gradient of
\begin{equation}
\partial_x p_{rel} ~=~ \mp\frac{\Delta p_{21}}{2W} ~=~ \mp\frac{2\gamma\Delta R}{W^3}
\end{equation}
relative to the common mean internal pressure (with negative sign for $D_1$ and positive sign for $D_2$).

The pressure gradient drives a flow $u_{rel}$ in $x$-direction, relative to the common mean flow $U$, with
\begin{equation}
u_{rel}~=~-0.035\,\frac{WH}{2\mu_d}\cdot\partial_x p_{rel}
 ~=~ \pm0.029\cdot\frac{1}{\lambda Ca}\frac{\Delta R}{W}\cdot U,\label{eq:urel}
\end{equation}
(with positive sign for $D_1$ and negative sign for $D_2$) with the velocity-pressure-relation \eqref{eq:vel_pressure}, the channel aspect ratio $W/H=1.2$. Only small deviations from this relation are expected due to the opening angle, aspect ratio of the channels and other effects. The velocity profile in Figure \ref{fig:pressure_contour}b shows the relative flow in each half of the channel, superimposed with the mean flow of strength $U$.

\section{Restricting the Gap Width in Simulations}\label{sec:appendix_gap_width}
To avoid numerical inaccuracy due to diverging Green's functions in the limit of small gaps between the droplet interfaces and channel walls, the numerical scheme enforces a lower limit $w_\textrm{min}$ on the width of the gap. Validations have shown robust behavior of the numerical code at gap widths of $w_\textrm{min}=10^{-2}H$ and were able to produce results down to $w_\textrm{min}=10^{-3}H$, with $H$ the channel height.

\begin{figure}
\centering
\includegraphics[width=\textwidth]{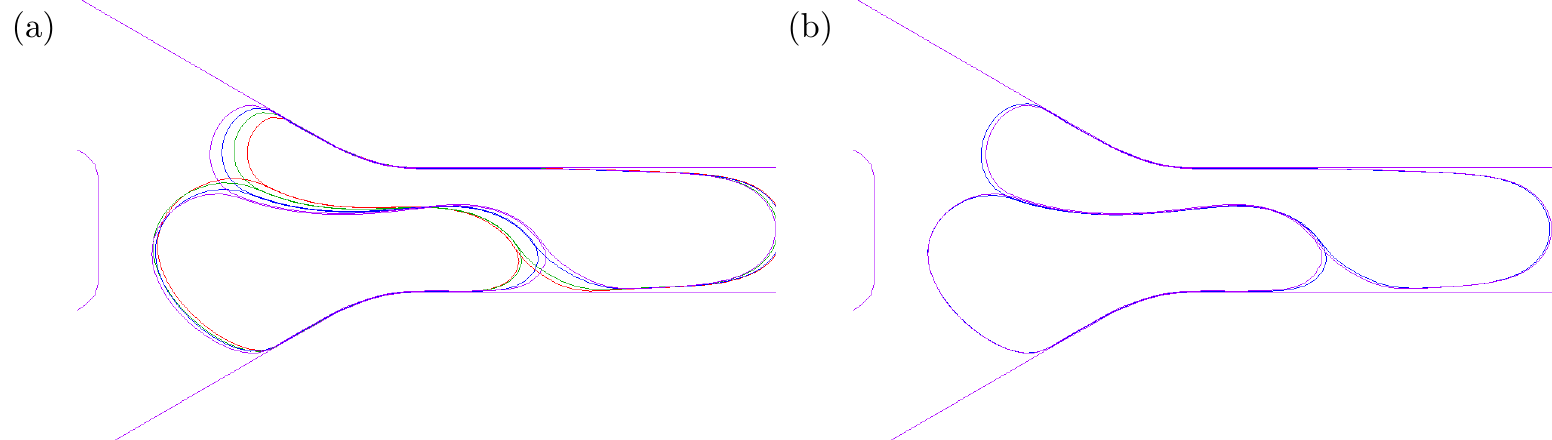}
\caption{Impact of restricting the gap width between droplet interfaces and channel walls.
\subfigure{a} Droplet shape at gap width $w_\textrm{min}=0.01H$ (purple), $w_\textrm{min}=0.005H$ (blue), $w_\textrm{min}=0.002H$ (green), $w_\textrm{min}=0.001H$ (red), for constant droplet size $a=1.828$. At larger $w_\textrm{min}$, the droplets appear longer.
\subfigure{b} Droplet shape at gap width $w_\textrm{min}=0.01H$ and size $a=1.828$ (purple), $w_\textrm{min}=0.005H$ and $a=1.84$ (blue). The change in gap width can be compensated by a change in droplet size to obtain the same behavior.
All simulations at $Ca=0.06$, $\lambda=0.8$, $\delta_0=0.01H$, $\alpha=30^\circ$, after approximately 3,500 time steps at 25,000 degrees of freedom per simulation.
}
\label{fig:appendix_gapwidth}
\end{figure}

Varying the gap width $w_\textrm{min}$ for fixed channel geometry will modify the effective confinement (the distance of the walls from the droplet slightly changes) so that a droplet of constant volume will appear more elongated when the gap width is increased (Figure \ref{fig:appendix_gapwidth}a). Likewise a droplet of prescribed volume with lower gap width will appear shorter in length, as the volume is conserved while the lateral interfaces move closer to the side walls.

We carry out two sets of numerical computations with varying numerical gap widths. One in which the droplet volume is fixed, and one in which we adapt the volume to compensate for the modified gaps. The droplets are initiated upstream in the channel and numerically followed as they move into the junction. Each simulation has approximately 25,000 degrees of freedom and is integrated over a total of 3,500 time steps. 

Figure \ref{fig:appendix_gapwidth}a shows the droplet shape for 4 different gap widths and a constant droplet volume. The shapes are very similar, indicating a similar breakup process, but the change in length due to the varying gap width is apparent. In a second set of simulations we compensated the effect of different gaps by increasing the droplet volume as the gap width is reduced. With $A$ the total surface area of the droplet and $\Delta w_\textrm{min}$ the change in minimum gap width, we change the volume by $\Delta V=-A\Delta w_\textrm{min}$. For a 5-fold decrease from $w_\textrm{min}=0.005$ (blue) to $w_\textrm{min}=0.001$ (red), a 0.7\% increase in droplet size (a 2\% increase in volume) fully compensates the gap effect. This is shown in Figure \ref{fig:appendix_gapwidth}b where the droplet shapes are almost indistinguishable despite a two-fold variation of the gap width and after a forward time integration over 3,500 steps. Consequently, the results only very slightly depend on the value of the numerically enforced gap width. Our numerical scheme thus faithfully captures the flow.

\end{document}